\documentclass[conference]{IEEEtran}
\IEEEoverridecommandlockouts
\usepackage{preamble}

\begin{document}

\title{Proposal on Model Based Current Overshoot Suppression of Receiver Side Coil in Drone Wireless Power Transfer System
}
\author{\IEEEauthorblockN{Kota Fujimoto}
\IEEEauthorblockA{\textit{Graduate School of Frontier Sciences} \\
\textit{The University of Tokyo}\\
Kashiwa, Japan \\
fujimoto.kota21@ae.k.u-tokyo.ac.jp}
\and
\IEEEauthorblockN{Takumi Hamada}
\IEEEauthorblockA{\textit{Graduate School of Frontier Sciences} \\
\textit{The University of Tokyo}\\
Kashiwa, Japan \\
hamada.takumi21@ae.k.u-tokyo.ac.jp}
\and
\IEEEauthorblockN{Hiroshi Fujimoto}
\IEEEauthorblockA{\textit{Graduate School of Frontier Sciences} \\
\textit{The University of Tokyo}\\
Kashiwa, Japan \\
fujimoto@k.u-tokyo.ac.jp}
}

\maketitle
\thispagestyle{titlepagestyle} 

\begin{abstract}
  This paper proposes a model-based control method in the wireless power transfer (WPT) system by operating a semi-bridgeless active rectifier (SBAR) to suppress the secondary coil current overshoot.
  By damping the current overshoot, it is possible to reduce the rectifier's rated current and decrease the rectifier's size, which is beneficial for the lightweight-oriented system such as drones.
  In the control method, an inverse of the plant model is used to calculate the reference input to the system.
  The current overshoot is reduced by operating the SBAR under the duty ratio calculated from the model.
  To confirm the performance of the proposed method, the simulation and the experiment using the WPT prototype are conducted.
  The experimental results show that the proposed method can suppress the secondary coil current overshoot.
  The results suggest it is possible to realize the lighter secondary system by applying the proposed method.
\end{abstract}

\begin{IEEEkeywords}
  wireless power transfer, semi-bridgeless active rectifier, overshoot suppression, current control
\end{IEEEkeywords}

\section{Introduction}
\begin{figure}[t]
    \centering
    \includegraphics[width=0.74\linewidth]{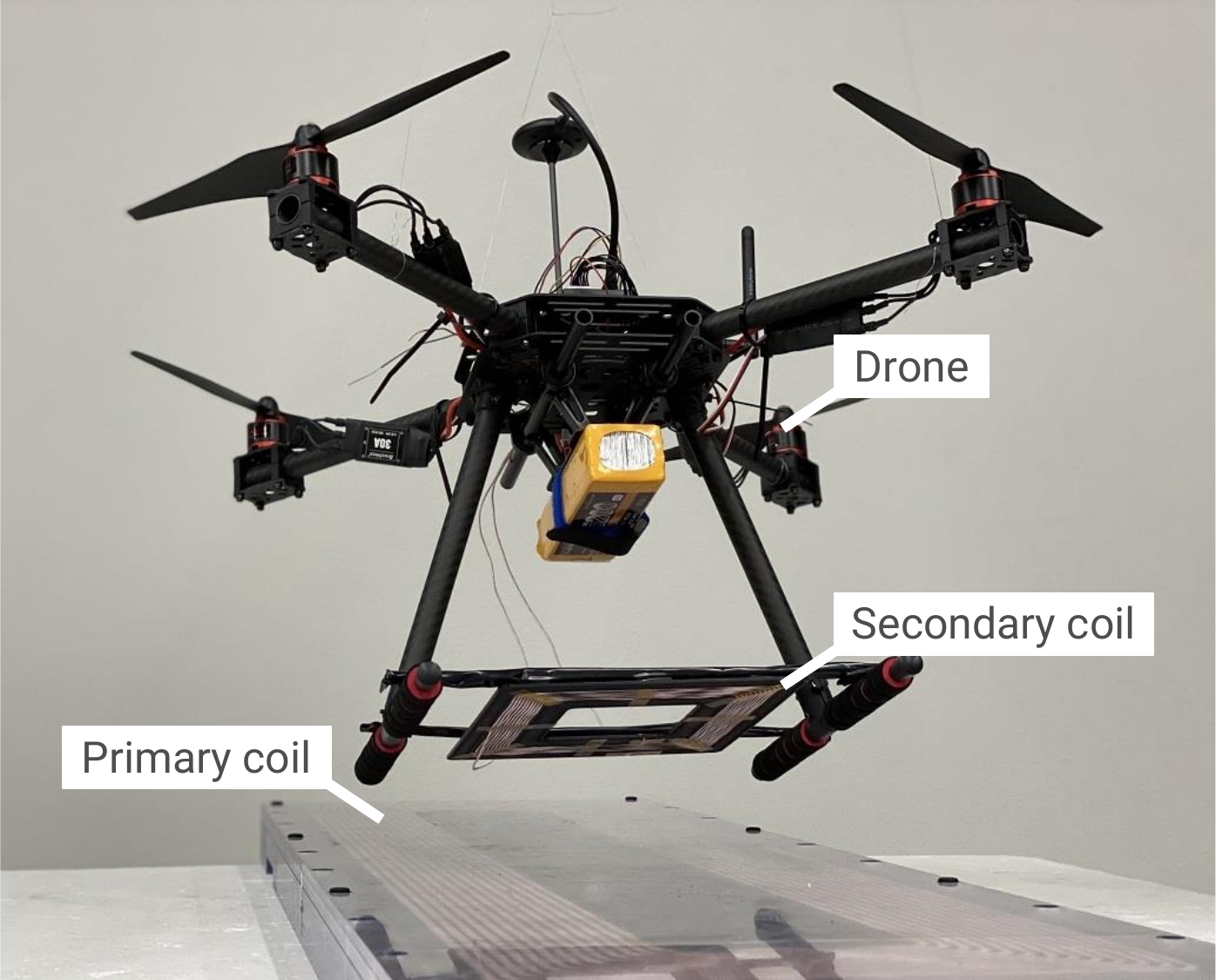}
    \caption{Diagram of the drone in-flight charging system.}
    \label{fig: diagram of drone in-flight charging}
  \end{figure}

  There are some studies that aim to realize the wireless power transfer (WPT) system for a flying drone.
  The feasibility of applying the WPT system to a flying drone is verified in \cite{Brown1969} to extend the flight duration, in which the microwave beam is implemented as an energy medium. 
  The main bottleneck for the microwave WPT system for drones is low efficiency, and recently there have been several studies that have tried to improve the system efficiency. 
  In \cite{Takahashi2020}, the port-to-port efficiency is $\SI{31.4}{\percent}$, which is not practically sufficient for drones. 
  On the other hand, some studies show that a magnetic resonant circuit helps realize the WPT system for flying drones\cite{arteaga2019,Campi2019,zhang2022,Shao2022}.
  In \cite{arteaga2019}, the average link efficiency, which is the same role as a port-to-port efficiency, is approximately $\SI{90}{\percent}$ despite the dynamic conditions of flying drones. 
  Although they both have their advantages, it is generally said that the magnetic resonant WPT system is more efficient than the microwave WPT system.
  \cref{fig: diagram of drone in-flight charging} shows a diagram of the drone wireless in-flight charging system with the magnetic resonant WPT system.
  To realize this system, it is needed to make the secondary system as light as possible.

  \par
\begin{figure}[t]
    \centering
    \subfigure[]{
      \includegraphics[width=0.35\linewidth]{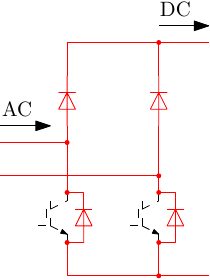}
      \label{fig: circuit diagram of SBAR with rectification mode}}
    \subfigure[]{
      \includegraphics[width=0.35\linewidth]{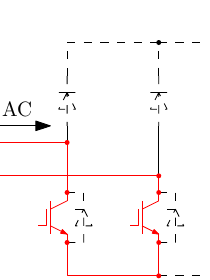}
      \label{fig: circuit diagram of SBAR with short mode}}
    \caption{Circuit diagram of the SBAR. (a) Rectification mode (b) Short mode}
    \label{fig: Operation circuit diagram of the SBAR at the secondary side}
  \end{figure}

  A magnetic resonant WPT system has a rectifier on the secondary side.
  In the drone wireless in-flight charging system, it could happen that excessive current flows into the battery due to a sudden change of the mutual inductance.
  Therefore, it is necessary to choose a lighter and a power-controllable rectifying system to realize the in-flight charging system.
  There are several types of rectifiers, such as a full-bridge diode rectifier\cite{Babaki2021}, a full-bridge active rectifier with semiconductor switches\cite{lovison2019,Okada2021}, and a semi-bridgeless active rectifier (SBAR)\cite{Colak2015,Zhao2017,Nagai2021}, which has two diodes and two semiconductor switches as shown in \cref{fig: Operation circuit diagram of the SBAR at the secondary side}.
  Using a full-bridge diode rectifier, it is unavoidable to load a DC-DC converter to control the power to a battery, which increases the weight of drones.
  Although it is possible to control the load power by using a full-bridge active rectifier without a DC-DC converter, the system becomes more complex and expensive than an SBAR.
  On the other hand, an SBAR does not need a DC-DC converter, and it is composed of only two semiconductor switches.
  Therefore, an SBAR is the most suitable for a drone WPT system in terms of weight and simplicity.
  \par
  The SBAR often operates with an unsynchronized ON-OFF switching method\cite{Gunji2015}.
  This method can be easily implemented as it does not need an alternative current sensor.
  The operation circuit diagram of the SBAR is shown in \cref{fig: Operation circuit diagram of the SBAR at the secondary side}.
  The transferred secondary AC current is rectified at the SBAR with rectification mode (RM) to charge the battery as shown in \cref{fig: circuit diagram of SBAR with rectification mode}.
  When the battery is fully charged, the SBAR operation mode is switched to short mode (SM) as shown in \cref{fig: circuit diagram of SBAR with short mode}, so that the current flow to the battery is cut off.
  By changing the operation mode, it is possible to prevent the battery from overcharging.
  However, when the mode switches from the RM to the SM, the secondary coil current overshoot occurs.
  There have been no studies addressing this problem.
  To avoid raising the rated current and increasing the size of the rectifier, we focus on the overshoot suppression in the SBAR system with an unsynchronized ON-OFF switching method.

  \begin{figure}[t]
    \centering
    \includegraphics[width=1.0\linewidth]{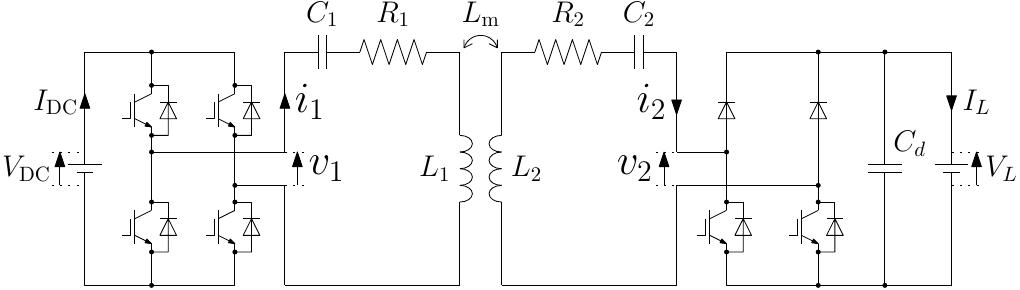}
    \caption{Circuit diagram of the magnetic resonant WPT system with the SBAR.}
    \label{fig: circuit diagram of magnetic resonant WPT with SBAR}
  \end{figure}

  \par
  This paper focuses on the secondary coil current control to resolve the above problem.
  By switching between the RM and the SM alternately and gradually extending the SM period, it is possible to suppress the coil current overshoot.
  The model of the WPT system derived in this paper determines the ratio of the SM period to the operation period.
  The experimental results show that the overshoot is suppressed by applying the control method to the SBAR system.

\section{Topology analysis}

  \begin{figure}[t]
    \centering
    \includegraphics[width=0.8\linewidth]{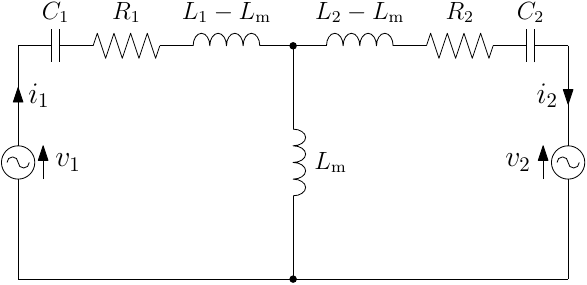}
    \caption{Equivalent circuit model of the WPT part.}
    \label{fig: Equivalent circuit model of the system}
  \end{figure}

  \begin{figure}[t]
    \centering
    \includegraphics[width=0.8\linewidth]{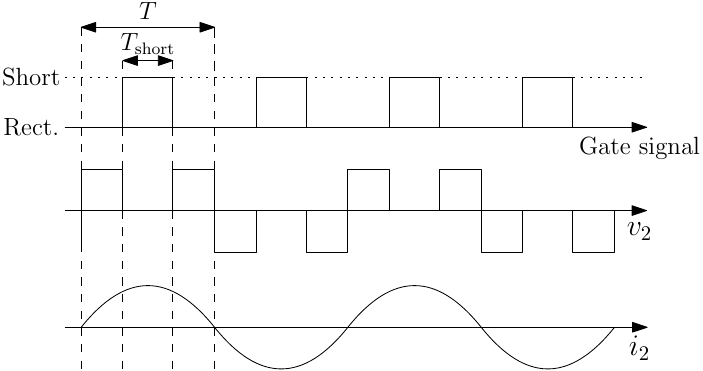}
    \caption{Ideal operating waveforms of the SBAR.}
    \label{fig: Switching scheme of the SBAR}
  \end{figure}

  \begin{figure*}[t]
    \centering
    \includegraphics[width=0.65\linewidth]{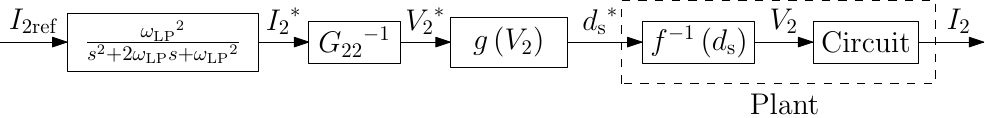}
    \caption{Block diagram of the control system.}
    \label{fig: block diagram of the control system}
  \end{figure*}

  The circuit used in this paper is shown in \cref{fig: circuit diagram of magnetic resonant WPT with SBAR}.
  The circuit consists of a direct current (DC) voltage source, full-bridge inverter, series-series compensated resonant circuit, SBAR, and smoothing capacitor.
  $v$ and $i$ are the instantaneous value of the voltage and current, $V$ and $I$ are the fundamental amplitude of $v$ and $i$, $L$, $C$, and $R$ are self-inductance, resonant capacitor, and internal resistor, respectively.
  The subscript 1, 2 represent the value of the primary side, the secondary side, respectively.
  $V_{\mathrm{DC}}$, $I_{\mathrm{DC}}$ and $V_{\mathrm{L}}$, $I_{\mathrm{L}}$ are the value of the primary DC voltage and current, and the secondary DC voltage and current.
  \cref{fig: Equivalent circuit model of the system} is the equivalent circuit of the system shown in \cref{fig: circuit diagram of magnetic resonant WPT with SBAR}.
  The circuit equation is expressed as follows:
  \begin{subequations}
    \label{eq: the circuit equation}
    \begin{align}
      v_1 &= R_1 i_1 + \frac{1}{C_1} \int i_1 \mathrm{dt} + L_1 \frac{\mathrm{d}i_1}{\mathrm{d}t} - L_{\mathrm{m}} \frac{\mathrm{d}i_2}{\mathrm{d}t},
      \label{eq: the circuit equation of the primary circuit} \\
      v_2 &= - R_2 i_2 - \frac{1}{C_2} \int i_2 \mathrm{dt} - L_2 \frac{\mathrm{d}i_2}{\mathrm{d}t} + L_{\mathrm{m}} \frac{\mathrm{d}i_1}{\mathrm{d}t}.
      \label{eq: the circuit equation of the secsondary circuit}
    \end{align}
  \end{subequations}
  Under the resonant condition, focusing on the fundamental component, the waveforms are expressed as
  \begin{subequations}
    \label{eq: the waveforms of the system}
    \begin{align}
      i_1 &= I_1(t) \sin{\omega t},
      \label{eq: the waveform of the primary current under the resonant condition} \\
      v_1 &= V_1(t) \sin{\omega t},
      \label{eq: the waveform of the primary voltage under the resonant condition} \\
      i_2 &= I_2(t) \cos{\omega t},
      \label{eq: the waveform of the secondary current under the resonant condition} \\
      v_2 &= V_2(t) \cos{\omega t}.
      \label{eq: the waveform of the secondary voltage under the resonant condition}
    \end{align}
  \end{subequations}
  $\omega$ is the angular frequency of the inverter, which equals to the resonant frequency of the primary and secondary circuit.
  By substituting \cref{eq: the waveforms of the system} into \cref{eq: the circuit equation}, the following equations are obtained:
  \begin{subequations}
    \label{eq: the obtained circuit equations}
    \begin{align}
      2L_1 \frac{\mathrm{d}I_1(t)}{\mathrm{d}t} + R_1 I_1 + \omega L_{\mathrm{m}} I_2 - V_1 &= 0,
      \label{eq: the obtained circuit equation of the primary circuit} \\
      2L_2 \frac{\mathrm{d}I_2(t)}{\mathrm{d}t} + R_2 I_2 - \omega L_{\mathrm{m}} I_1 - V_2 &= 0.
      \label{eq: the obtained circuit equation of the secondary circuit}
    \end{align}
  \end{subequations}
  These equations are derived under the approximation that the frequency of the current envelope is sufficiently smaller than the operating frequency.
  \par
  The Laplace form of the equations \cref{eq: the obtained circuit equations} are shown as
  \begin{subequations}
    \label{eq: the laplace forms of the circuit equations}
    \begin{align}
      2 L_1 s I_1(s) + R_1 I_1(s) + \omega L_{\mathrm{m}} I_2(s) + V_1(s) &= 0, \label{eq: the laplace form of the primary circuit equation} \\
      2 L_2 s I_2(s) + R_2 I_2(s) - \omega L_{\mathrm{m}} I_1(s) + V_2(s) &= 0. \label{eq: the laplace form of the secondary circuit equation}
    \end{align}
  \end{subequations}
  By solving \cref{eq: the laplace forms of the circuit equations}, $I_1(s)$ and $I_2(s)$ are expressed as
  \begin{subequations}
    \begin{align}
      I_1(s) = \frac{\left( \alpha_1 s + \beta_1 \right) V_1(s)}{s^2 + 2\zeta \omega_{\mathrm{n}} s + {\omega_{\mathrm{n}}}^2} - \frac{\gamma V_2(s)}{s^2 + 2\zeta \omega_{\mathrm{n}} s + {\omega_{\mathrm{n}}}^2}, \label{eq: the laplace form of the primary current} \\
      I_2(s) = \frac{\gamma V_1(s)}{s^2 + 2\zeta \omega_{\mathrm{n}} s + {\omega_{\mathrm{n}}}^2} - \frac{\left( \alpha_2 s + \beta_2 \right) V_2(s)}{s^2 + 2\zeta \omega_{\mathrm{n}} s + {\omega_{\mathrm{n}}}^2}, \label{eq: the laplace form of the secondary current}
    \end{align}
  \end{subequations}
  where $\zeta$, $\omega_{\mathrm{n}}$, $\alpha_1$, $\beta_1$, $\alpha_2$, $\beta_2$, and $\gamma$ are shown as
  \begin{equation}
    \begin{aligned}
      &\zeta = \frac{L_1 R_2 + L_2 R_1}{\sqrt{4 L_1 L_2}} \cdot \frac{1}{\sqrt{R_1 R_2 + \left( \omega L_{\mathrm{m}} \right)^2 }}, \\
      &\omega_{\mathrm{n}} = \frac{1}{2} \sqrt{\frac{R_1 R_2 + \left( \omega L_{\mathrm{m}} \right)^2}{L_1 L_2}}, \\
      &\alpha_1 = \frac{1}{2 L_1}, \hspace{5mm} \beta_1 = \frac{R_2}{4 L_1 L_2}, \\
      &\alpha_2 = \frac{1}{2 L_2}, \hspace{5mm} \beta_2 = \frac{R_1}{4 L_1 L_2}, \hspace{5mm} \gamma = \frac{\omega L_{\mathrm{m}}}{4 L_1 L_2}. \\
    \end{aligned}
  \end{equation}

\section{Control strategy for current overshoot suppression}

  \begin{table}[t]
    \centering
    \caption{Circuit parameters}
    \label{table: circuit parameters}
    \scalebox{0.85}{
      \begin{tabular}{ll} \toprule
        Parameter & Value \\ \midrule
        Primary inductance $L_1$ & $\SI{236}{\micro H}$ \\
        Secondary inductance $L_2$ & $\SI{18.9}{\micro H}$ \\
        Mutual inductance $L_{\mathrm{m}}$ & $\SI{6.25}{\micro H}$ \\
        Primary resistor $R_1$ & $\SI{108}{m \ohm}$ \\
        Secondary resistor $R_2$ & $\SI{32.5}{m \ohm}$ \\
        Input voltage $V_{\mathrm{DC}}$ & $\SI{30}{\volt}$ \\
        Load voltage $V_L$ & $\SI{30}{\volt}$ \\
        Resonance frequency $f$ & $\SI{85}{k\hertz}$ \\ \bottomrule
      \end{tabular}
    }
  \end{table}

  \begin{figure*}[t]
    \centering
    \subfigure[]{
      \includegraphics[width=0.25\linewidth]{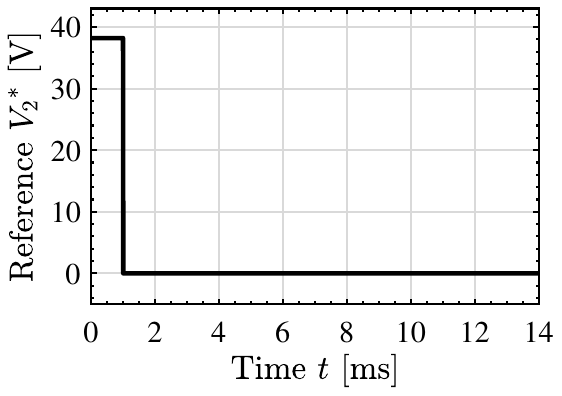}
      \label{fig: secondary voltage reference when is stepped}}
    \subfigure[]{
      \includegraphics[width=0.25\linewidth]{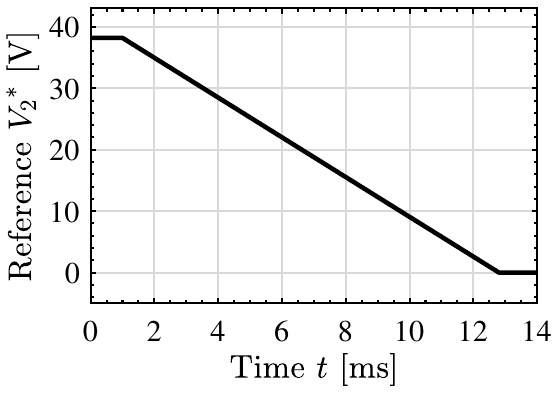}
      \label{fig: secondary voltage reference when is ramped}}
    \subfigure[]{
      \includegraphics[width=0.25\linewidth]{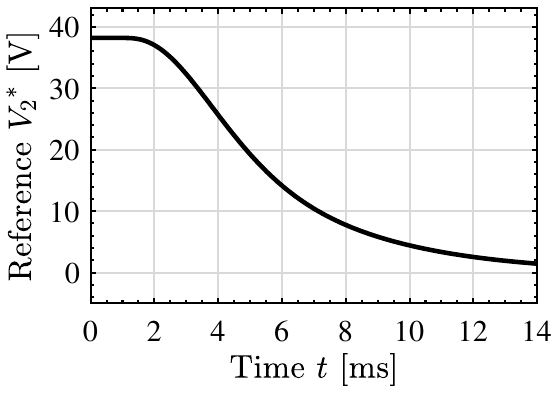}
      \label{fig: secondary voltage reference when is proposed}}
    \subfigure[]{
      \includegraphics[width=0.25\linewidth]{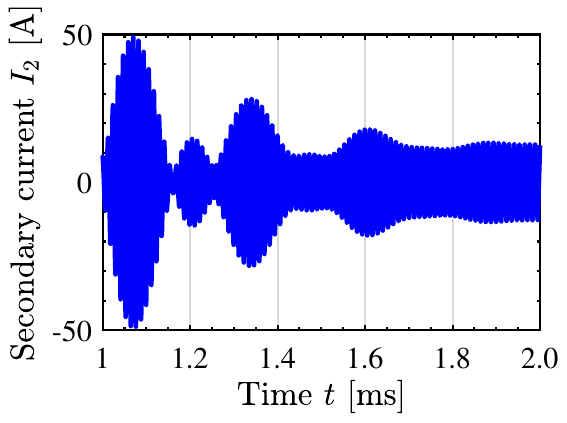}
      \label{fig: simulation results of the secondary current when is stepped}}
    \subfigure[]{
      \includegraphics[width=0.25\linewidth]{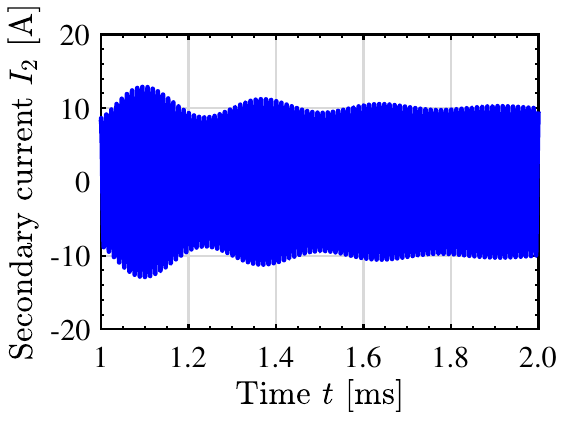}
      \label{fig: simulation results of the secondary current when is ramped}}
    \subfigure[]{
      \includegraphics[width=0.25\linewidth]{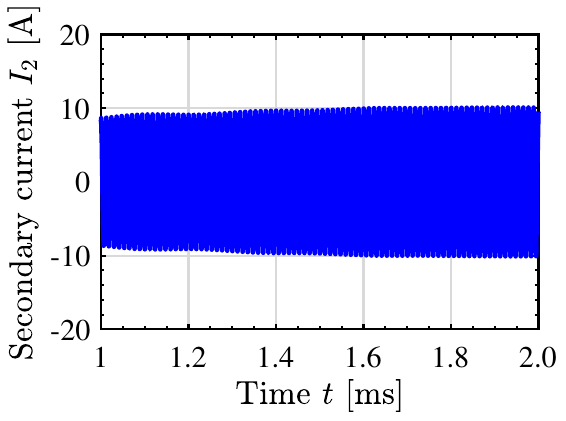}
      \label{fig: simulation results of the secondary current when is proposed}}
    \caption{Reference values of ${V_2}^{*}$ and simulation results of the secondary current. (a) Step-type reference (b) Ramp-type reference (c) Proposed reference (d) Result with the step-type reference (e) Result with the ramp-type reference (f) Result with the proposed reference}
    \label{fig: secondary voltage reference and simulation results of the secondary current}
  \end{figure*}

  This section describes how the SBAR is operated.
  \cref{fig: Switching scheme of the SBAR} shows the ideal operating waveforms of the SBAR.
  It shows the gate signal sent to the lower arm of the rectifier, the secondary voltage $v_2$, and the secondary current $i_2$, respectively.
  $T$ is the rectifier's operating period.
  In this paper, $T$ is defined as a half of the switching period of the primary inverter.
  $T_{\mathrm{short}}$ is the short period in $T$ in which the lower arm of the rectifier is on.
  During this period, the power is not transferred to the load side.
  The duty ratio of the rectifier $d_{\mathrm{short}}$ is defined as
  \begin{equation}
    \label{eq: duty ratio of the rectifier}
    d_{\mathrm{short}} = \frac{T_{\mathrm{short}}}{T}.
  \end{equation}
  By deciding $d_{\mathrm{short}}$, the operation of the rectifier is determined.
  \par
  According to \cref{eq: the laplace form of the secondary current}, the transfer function from $V_2$ to $I_2$ is expressed as follows:
  \begin{equation}
    \label{eq: transfer function from V_2 to I_2}
    G_{22} = \frac{I_2(s)}{V_2(s)} = - \frac{\alpha_2 s + \beta_2}{s^2 + 2\zeta \omega_{\mathrm{n}} s + {\omega_{\mathrm{n}}}^2}.
  \end{equation}

  When a step signal of $V_2$ is input to the system by switching the SBAR from the RM to the SM, the coil current overshoot occurs because $\zeta$ is too small in a typical magnetic resonant WPT system.
  In order to suppress the overshoot in the step response of $I_2$, this paper proposes a SBAR control method.
  By controlling $V_2$ appropriately, the overshoot can be suppressed.
  \par
  \cref{fig: block diagram of the control system} is the block diagram of the control system, which shows how the SBAR is operated.
  $I_{2\mathrm{ref}}$, ${I_2}^{*}$, ${V_2}^{*}$, and ${d_{\mathrm{short}}}^{*}$ are the secondary current's reference value, the secondary current's calculated value, the secondary voltage's calculated value, and the calculated value of the duty ratio of the rectifier, respectively.
  $f\left( d_{\mathrm{short}} \right)$ is the function showing the relation between $V_2$ and $d_{\mathrm{short}}$, which is defined as
  \begin{equation}
    \label{eq: function which shows the relation between duty ratio to V2}
    V_2 = f\left( d_{\mathrm{short}} \right) = \frac{4}{\pi} V_L \left\{ 1 - \sin{ \left( \frac{\pi}{2} d_{\mathrm{short}} \right) } \right\}.
  \end{equation}
  \cref{eq: function which shows the relation between duty ratio to V2} is derived as the difference between the fundamental wave amplitude of the square wave and the fundamental wave amplitude of the square wave when operating the rectifier with a phase-shift control.
  $f^{-1}\left( V_2 \right)$ is the inversion of $f\left( d_{\mathrm{short}} \right)$.
  In this paper, ${V_2}^{*}$ is calculated using the inversion system of $G_{22}$.
  ${G_{22}}^{-1}$ is the non-proper system, so the low-pass filter shown in \cref{fig: block diagram of the control system} is placed before ${G_{22}}^{-1}$.
  Applying this scheme to the operation, the proposed method, in which the secondary coil current overshoot at the transition phase from the RM to the SM is suppressed, is implemented.

\section{Simulation and experiment}

  In order to validate the effectiveness of the proposed method, the simulation is carried out at first.
  The parameter is listed in \cref{table: circuit parameters}.
  Figs.~\ref{fig: secondary voltage reference when is stepped}, \ref{fig: secondary voltage reference when is ramped}, and \ref{fig: secondary voltage reference when is proposed} show the reference values of ${V_2}^{*}$ considered in the simulation.
  Each figure shows a step-type reference value, a ramp-type reference value, and a proposed reference value calculated using the model as shown in \cref{fig: block diagram of the control system}, respectively.
  The slope of the ramp-type reference value is decided such that the time constant is the same as that of the proposed method.
  \par
  Figs.~\ref{fig: simulation results of the secondary current when is stepped}, \ref{fig: simulation results of the secondary current when is ramped}, and \ref{fig: simulation results of the secondary current when is proposed} show the simulation results.
  All the waveforms once converge at \SI{4}{\milli\second}.
  The convergence of the secondary coil current amplitude at \SI{4}{\milli\second} is \SI{10.5}{\ampere}.
  The waveforms after \SI{4}{\milli\second} are shown because the proposed reference value converges around \SI{14}{\milli\second}.
  \par
  \cref{fig: simulation results of the secondary current when is stepped} shows that the maximum current amplitude is tremendously large if there is no controller in the secondary system.
  \cref{fig: simulation results of the secondary current when is ramped} shows that the ramp-type reference trajectory cannot wholly suppress the overshoot.
  On the other hand, \cref{fig: simulation results of the secondary current when is proposed} shows that the proposed method can completely suppress the overshoot.
  It is because the inverse of the plant model is implemented in the control system, which is able to suppress the overshoot ideally.
  \cref{table: maximum value of the secondary coil current overshoot} shows all the maximum values of the secondary coil current between \SI{0}{\milli\second} and \SI{4}{\milli\second} in the simulation.
  According to \cref{table: maximum value of the secondary coil current overshoot}, it can be seen that the proposed method is numerically superior to the other methods.
  These simulation results verify the proposed method to dump the overshoot of the secondary coil current.

  \begin{table}[t]
    \centering
      \caption{Maximum values of the secondary coil current overshoot in the simulation}
      \label{table: maximum value of the secondary coil current overshoot}
      \scalebox{0.9}{
        \begin{tabular}{lll} \toprule
          ${V_2}^{*}$ trajectory & Maximum current amplitude & Overshoot \\ \midrule
          Step-type & $\SI{49.1}{\ampere}$ & \SI{368}{\percent} \\
          Ramp-type & $\SI{11.4}{\ampere}$ & \SI{8.6}{\percent} \\
          Proposed & $\SI{10.5}{\ampere}$ & \SI{0.0}{\percent} \\ \bottomrule
        \end{tabular}
      }
  \end{table}
  \par
\begin{figure}[t]
    \centering
      \includegraphics[width=0.8\linewidth]{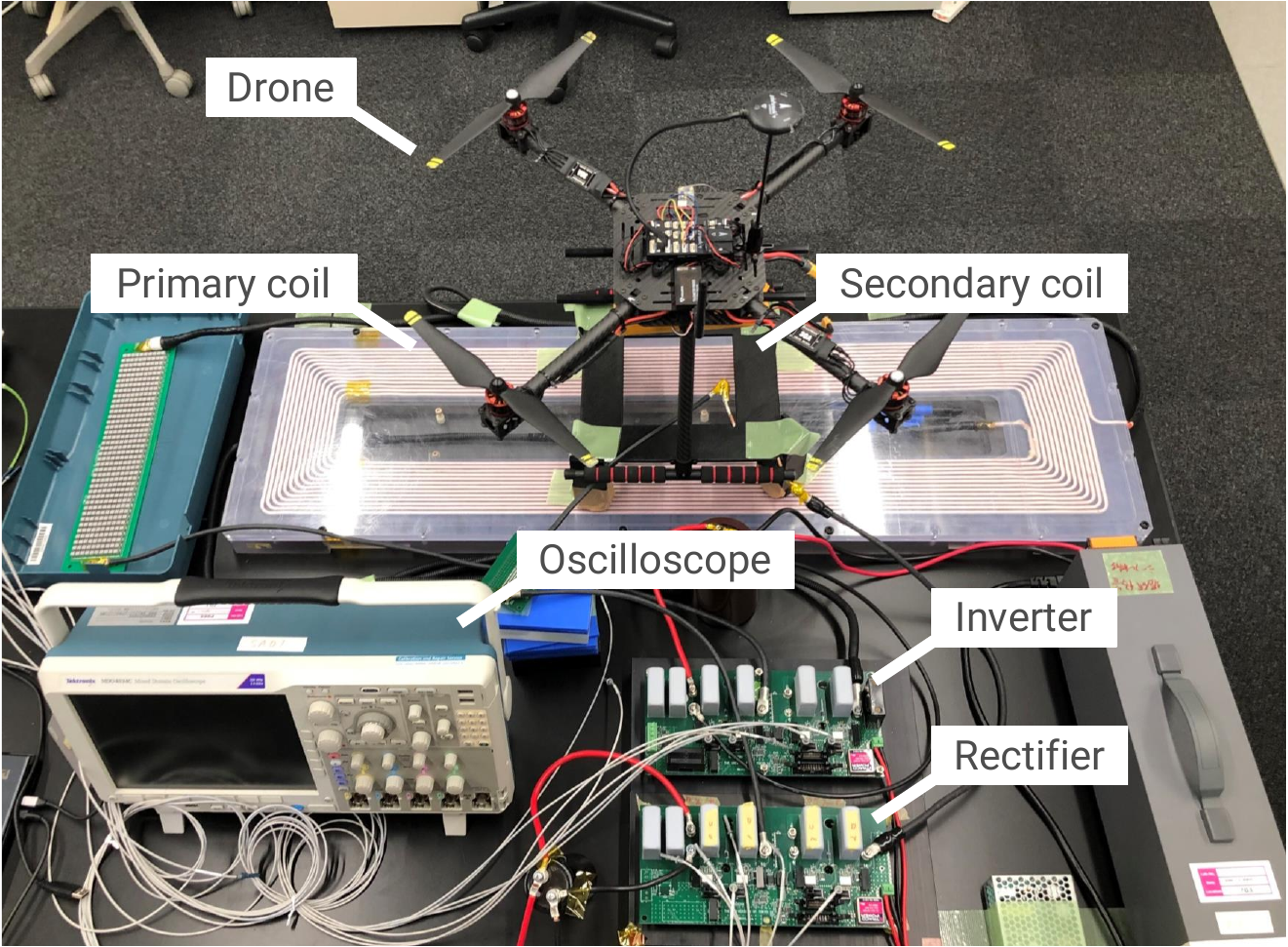}
      \caption{Experimental prototype of the system.}
    \label{fig: experimental prototype of the system}
  \end{figure}

  With the above simulation results, experiments are performed to verify the feasibility of the proposed method.
  \cref{fig: experimental prototype of the system} shows an experimental prototype setup based on the parameters listed in \cref{table: circuit parameters}.
  The calculation shown in Fig. 6 is implemented with a digital signal processing (DSP) controller.
  The reference values of ${V_2}^{*}$ are the same as that shown in \cref{fig: secondary voltage reference and simulation results of the secondary current}.
  The calculated value ${d_{\mathrm{short}}}^{*}$ is input to the lower arm of the rectifier as the gate signal.
  \par

\begin{figure}[t]
    \centering
    \subfigure[]{
      \includegraphics[width=0.45\linewidth]{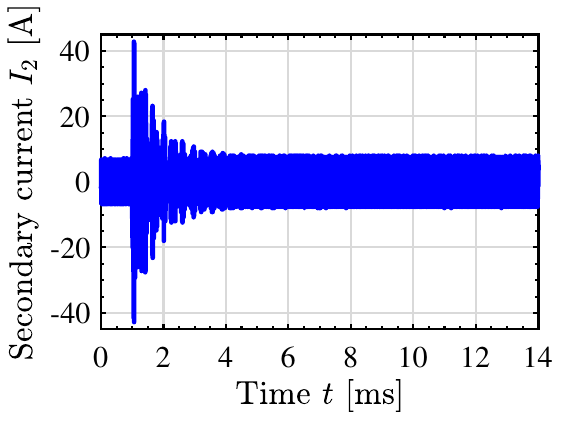}
      \label{fig: secondary current in experimental result when is stepped}}
    \subfigure[]{
      \includegraphics[width=0.45\linewidth]{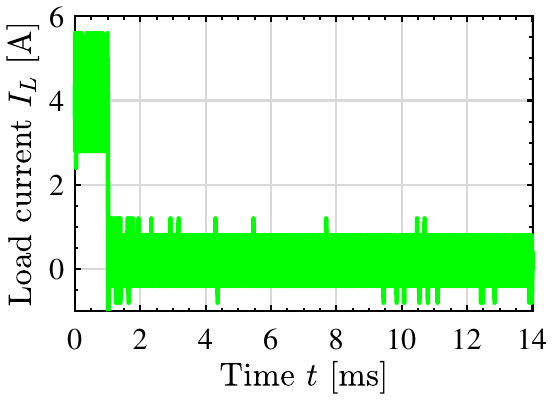}
      \label{fig: load current in experimental result when is stepped}}
    \caption{Experimental waveforms with the step-type reference value of ${V_2}^{*}$. (a) Secondary current (b) Load current}
    \label{fig: experimental result when is stepped}
  \end{figure}

\begin{figure}[t]
    \centering
    \subfigure[]{
      \includegraphics[width=0.45\linewidth]{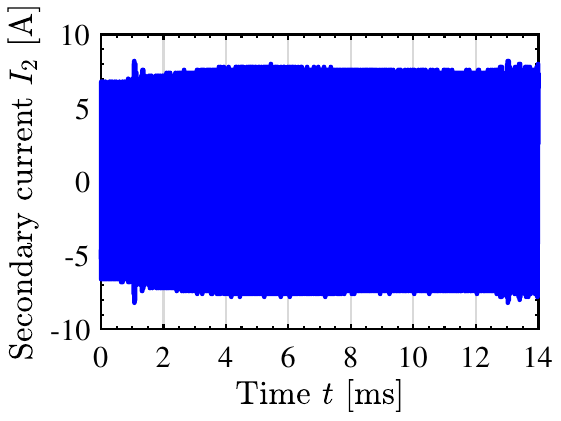}
      \label{fig: secondary current in experimental result when is ramped}}
    \subfigure[]{
      \includegraphics[width=0.45\linewidth]{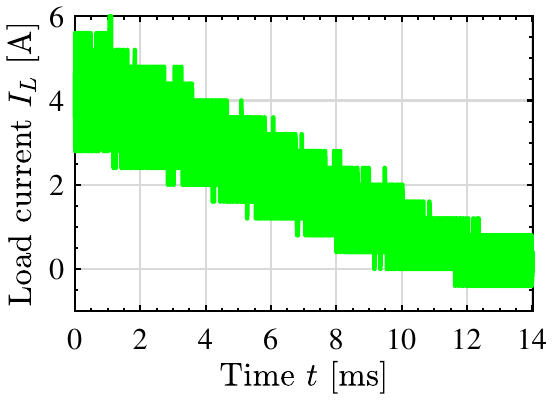}
      \label{fig: load current in experimental result when is ramped}}
    \caption{Experimental waveforms with the ramp-type reference value of ${V_2}^{*}$. (a) Secondary current (b) Load current}
    \label{fig: experimental result when is ramped}
  \end{figure}

\begin{figure}[t]
    \centering
    \subfigure[]{
      \includegraphics[width=0.45\linewidth]{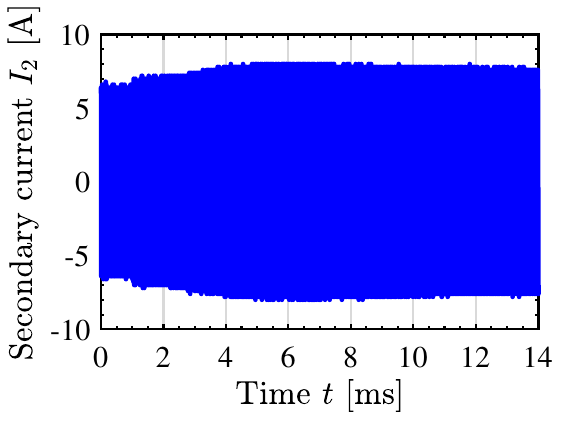}
      \label{fig: secondary current in experimental result when is made with proposed method}}
    \subfigure[]{
      \includegraphics[width=0.45\linewidth]{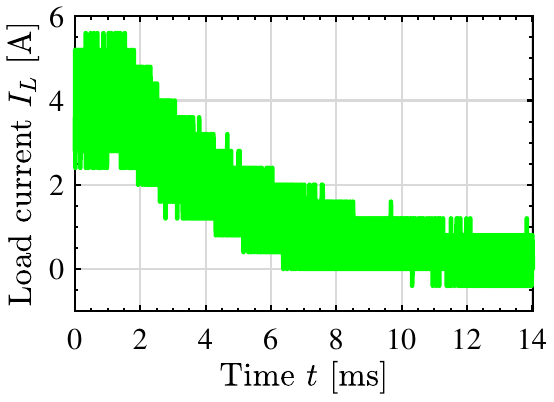}
      \label{fig: load current in experimental result when is made with proposed method}}
    \caption{Experimental waveforms with the proposed reference value of ${V_2}^{*}$. (a) Secondary current (b) Load current}
    \label{fig: experimental result when is made with proposed method}
  \end{figure}

  Figs. \ref{fig: experimental result when is stepped}--\ref{fig: experimental result when is made with proposed method} show the experimental results.
  Figs. \ref{fig: secondary current in experimental result when is stepped}, \ref{fig: secondary current in experimental result when is ramped}, and \ref{fig: secondary current in experimental result when is made with proposed method} show the waveforms of the secondary current.
  Their tendencies  are similar to the simulation results as shown in \cref{fig: secondary voltage reference and simulation results of the secondary current}, which shows the feasibility of the proposed method.
  \par
  Figs. \ref{fig: load current in experimental result when is stepped}, \ref{fig: load current in experimental result when is ramped}, and \ref{fig: load current in experimental result when is made with proposed method} show the waveforms of the load current.
  It is observed that the load current flows into the battery in Figs. \ref{fig: load current in experimental result when is ramped} and \ref{fig: load current in experimental result when is made with proposed method} after the switching mode of the rectifier is changed from the RM to the SM.
  In terms of the transferred power control, it can be said that the step method is more accessible than the other methods; however, it is possible to control the energy including the power transferred after changing the mode of the rectifier.
  In addition, there is no surge current in \cref{fig: load current in experimental result when is made with proposed method}, which means the proposed method has no adverse effect on the battery safety.
  These results suggest that the proposed method is valid to suppress the coil current overshoot and control the power transferred to the battery with the SBAR system, which decreases the rated current of the rectifier so that the lighter drone system is realized.
  \par
\begin{figure}[t]
    \centering
    \subfigure[]{
      \includegraphics[width=0.45\linewidth]{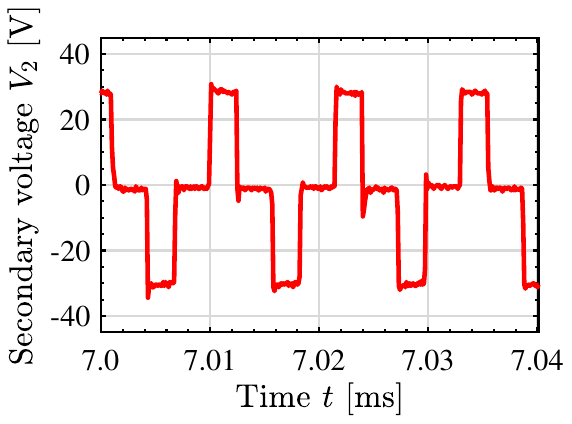}
      \label{fig: enlarged secondary voltage in experimental result when is made with proposed method}}
    \subfigure[]{
      \includegraphics[width=0.45\linewidth]{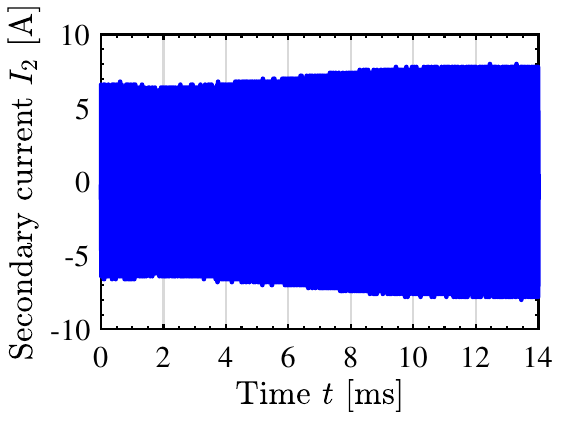}
      \label{fig: a worse case of the secondary current in experimental result when is made with proposed method}}
    \caption{Experimental waveforms with the proposed reference value of ${V_2}^{*}$ and the phases of the gate signal and the secondary current are shifted by $\frac{\pi}{2}$ from the state as shown in \cref{fig: Switching scheme of the SBAR}. (a) Enlarged secondary voltage (b) Secondary current}
    \label{fig: a worse case of the experimental results when is made with proposed method}
  \end{figure}

  On the other hand, it is needed to consider the case when the phase of the gate signal shifts from the state as shown in \cref{fig: Switching scheme of the SBAR} because the gate signal is not synchronized with the secondary current.
  According to \cref{fig: a worse case of the experimental results when is made with proposed method}, the coil current overshoot can be suppressed even if the phases of the gate signal and the secondary current are shifted by $\frac{\pi}{2}$ from the state as shown in \cref{fig: Switching scheme of the SBAR}.
  This result suggests that \cref{eq: function which shows the relation between duty ratio to V2} is robust to the fluctuation of the model which assumes that the phases of the gate signal and the secondary current match as shown in \cref{fig: Switching scheme of the SBAR}.

\section{Conclusion}
  In this paper, we proposed a novel model-based current overshoot suppression method with a two-mode operation SBAR.
  The control strategy is validated with the simulation and the experiment in which the maximum values of the secondary coil current overshoot are evaluated.
  The proposed method accomplishes a suppression of the secondary coil current overshoot, which leads to decreasing the rated current of the rectifier.
  By applying this method to the drone wireless in-flight charging system, the lighter secondary system is realized.
  \par
  Meanwhile, it should be noted that the proposed method assumes the perfect resonance condition.
  The robustness to the parameter fluctuation will be improved in future studies.

\section{Acknowledgment}
  This work was partly supported by JST-Mirai Program Grant Number JPMJMI21E2, JSPS KAKENHI Grant Number JP18H03768, and the New Energy and Industrial Technology Development Organization (NEDO) Project Number JPNP21005, Japan.



\bibliographystyle{IEEEtran}
\bibliography{list.bib}



\end{document}